\documentclass[12pt]{article}

\begin{document}

\begin{center}
{\Large\bf{} Application of the 3-space approach to the Bianchi II cosmological model}\\
Lau Loi So
\end{center}

\begin{abstract}
Einstein used 4-dimensional space time geometry to explain
gravity. However, in 1962, Baierlein, Sharp and Wheeler proposed a
Jacobi type timeless Lagrangian based on the 3-dimensional
geometry of space to reproduce the same physics.  In 2002, Barbour
$et$. $al$. further extended this idea and they call it 3-space
approach. Here we use Bianchi II cosmological model to demonstrate
the 3-space idea.  Indeed, we find that this theory is more
fundamental and the manipulation is more practical.  We recover
the known and find a new solutions.
\end{abstract}

\section{Introduction}
Almost century ago, Einstein used 4-dimensional space time
geometry to explain gravity. However, it seems that 4D concept is
not the most basic. As Dirac~\cite{Dirac} pointed out in 1958: `I
am inclined to believe from this that four-dimensional symmetry is
not a fundamental property of the physical world.' Four years
later, in 1962, Baierlein, Sharp and Wheeler~\cite{BSW} proposed a
Jacobi type timeless Lagrangian, i.e., BSW action. They laid down
a very nice and concrete foundation, but it did not attract much
attention for 40 years. Until recently in 2002, Barbour, Foster
and $\acute{\mbox{O}}$ Murchadha~\cite{RWR} extended this method
and they call it 3-space approach.  This theory uses 3-space,
without time, to describe the same physics as the 4-metric does.
The motivation of our work is not finding the Einstein solution
for the Bianchi II cosmological model.  But, in order to
appreciate the 3-space theory, we apply it to this universe model
for a simple testing. As a consequence, we have recovered the
known solution in 4-metric~\cite{Singh}, and also find a new set
of solutions. Moreover, we remark that finding these solutions
using 3-space theory is more practical.

\section{Einstein equations of motion in Bianchi II model}
The line element of the Bianchi II cosmological
model~\cite{metricII} in 4-spacetime is
\begin{equation}
ds^{2}=-dt^{2}+f_{1}^{2}dx^{2}+f_{2}^{2}dy^{2}-2xf_{2}^{2}dydz+(f_{1}^{2}+x^{2}f_{2}^{2})dz^{2},
\end{equation}
where $f_{1}$ and $f_{2}$ depend only on time $t$.  The metric
signature we use is $+2$, Latin indices indicate spatial and Greek
means spacetime. Solving the Einstein equation:
$G_{\mu\nu}=R_{\mu\nu}-\frac{1}{2}g_{\mu\nu}R$, the equations of
motion are
\begin{eqnarray}
&&G_{00}=\frac{\dot{f}_{1}\dot{f}_{1}}{f_{1}f_{1}}
+\frac{2\dot{f_{1}}\dot{f}_{2}}{f_{1}f_{2}}-\frac{f^{2}_{2}}{4f^{4}_{1}},\quad{}
G_{11}=-f^{2}_{1}\left(\frac{\ddot{f}_{1}}{f_{1}}+\frac{\ddot{f}_{2}}{f_{2}}
+\frac{\dot{f}_{1}\dot{f}_{2}}{f_{1}f_{2}}+\frac{f^{2}_{2}}{4f^{4}_{1}}\right),\label{3aSep2012}\\
&&xG_{22}=-xf^{2}_{2}\left(\frac{2\ddot{f}_{1}}{f_{1}}
+\frac{\dot{f}_{1}\dot{f}_{1}}{f_{1}f_{1}}-\frac{3f^{2}_{2}}{4f^{4}_{1}}\right)=-G_{23},\quad{}
G_{33}=G_{11}+x^{2}G_{22},
\end{eqnarray}
where $\dot{f}$ denotes differentiate w.r.t. time $t$. Confining
source free for Einstein equation in vacuum, we find that there
are only three independent equations of motion. These solutions
are known~\cite{Singh} by a simple transformation
$dt=f^{2}_{1}(T)f_{2}(T)dT$
\begin{equation}
f_{1}=a_{1}\sqrt{e^{T}\cosh{}T},\quad{}f_{2}=a_{2}/\sqrt{\cosh{}T},\label{29aAug2012}
\end{equation}
where $a_{1}, a_{2}$ are arbitrary constants and $T$ is another
time parameter.  After a simple checking, we find that
$a_{2}=\pm1$.

As there are only two unknown functions $f_{1}$ and $f_{2}$ in the
line element, why we have 3 equations from $G_{\mu\nu}$?  One
might speculate that it is overdetermined. However, it is not.
Here we give three examples to demonstrate this statement. (i)
Suppose $(dt/dT,f_{1},f_{2})=(1,\sqrt{t},1)$.  We check that this
trial solution satisfy $G_{00}$ and $G_{11}$, but fail to fulfill
$G_{22}$ requirement.  (ii) Suppose
$(dt/dT,f_{1},f_{2})=(e^{2T},e^{T},2)$. We find that they satisfy
$G_{00}$ and $G_{22}$, but violate $G_{11}$. (iii) Suppose
$(dt/dT,f_{1}f_{2})=(1,t,\sqrt{12}\,t)$.  This allows $G_{00}$
vanishes, but $G_{11}$ and $G_{22}$ cannot.  Hence $G_{00}$,
$G_{11}$ and $G_{22}$ are independent. Moreover, through these
three concrete examples, one might deduce that constant $f_{2}$ is
not adequate referring to the first two cases and we examine
$G_{\mu\nu}$ that indeed it is forbidden. This argument becomes
manifest by using the 3-space method in section 3 (i.e., see
(\ref{9aMar2012})).

\section{The 3-space approach for Bianchi II universe}
Here come to the 3-space approach for Bianchi II cosmological
model. The BSW type action~\cite{BSW,RWR} has the form
\begin{eqnarray}
I=\int{}d\lambda\int\sqrt{g}\sqrt{{}^{3}R}\sqrt{\cal{T}}\,d^{3}x,
\end{eqnarray}
where $\lambda$ is a parameter, the determinant
$\sqrt{g}=f^{2}_{1}(\lambda)f_{2}(\lambda)$, the three dimensional
scalar curvature $^{3}R=-f^{2}_{2}/(2f^{4}_{1})$, and `kinetic
energy'
\begin{eqnarray}
{\cal{T}}&=&G^{abcd}(\acute{g}_{ab}-\pounds_{\xi}g_{ab})(\acute{g}_{cd}-\pounds_{\xi}g_{cd})\nonumber\\
&=&-8\left(\frac{\acute{f}_{1}\acute{f}_{1}}{f_{1}f_{1}}+\frac{2\acute{f}_{1}\acute{f}_{2}}{f_{1}f_{2}}\right).
\end{eqnarray}
Note that $G^{abcd}=g^{ac}g^{bd}-g^{ab}g^{cd}$ is the DeWitt
supermetric, $\pounds$ is the Lie derivative, $\xi$ is the space
of the fields, $\acute{g}_{ab}$ and $\acute{f}$ mean differentiate
w.r.t. $\lambda$. The equations of motion can be obtained through
the Euler-Lagrange equation
\begin{eqnarray}
\frac{\partial}{\partial\lambda}\frac{\partial{\cal{L}}}{\partial\acute{g}_{ij}}
-\frac{\delta{\cal{L}}}{\delta{}g_{ij}}=0,
\end{eqnarray}
where the Lagrangian density ${\cal{L}}=\sqrt{^{3}Rg{\cal{T}}}$.
The momentum $p^{ij}=\partial{\cal{L}}/\partial\acute{g}_{ij}$ and
its corresponding non-vanishing components are
\begin{eqnarray}
p^{11}=-\frac{\sqrt{g}g^{11}}{N}\left(\frac{\acute{f}_{1}}{f_{1}}+\frac{\acute{f}_{2}}{f_{2}}\right),
&&p^{22}=-\frac{\sqrt{g}}{N}\left[g^{22}
\left(\frac{\acute{f}_{1}}{f_{1}}+\frac{\acute{f}_{2}}{f_{2}}\right)
+\frac{1}{f^{2}_{2}}\left(\frac{\acute{f}_{1}}{f_{1}}-\frac{\acute{f}_{2}}{f_{2}}\right)\right],\\
p^{23}=-\frac{\sqrt{g}g^{23}}{N}\left(\frac{\acute{f}_{1}}{f_{1}}+\frac{\acute{f}_{2}}{f_{2}}\right),
&&p^{33}=-\frac{\sqrt{g}g^{33}}{N}\left(\frac{\acute{f}_{1}}{f_{1}}+\frac{\acute{f}_{2}}{f_{2}}\right).
\end{eqnarray}
The scalar momentum is
\begin{eqnarray}
p=-\frac{2\sqrt{g}}{N}\left(\frac{2\acute{f}_{1}}{f_{1}}+\frac{\acute{f}_{2}}{f_{2}}\right),
\end{eqnarray}
where the lapse
$N:=\sqrt{{\cal{T}}/[4(^{3}R)]}=dt/d\lambda$~\cite{RWR} and the
associate expansion of $N$ is
\begin{equation}
0=\frac{\acute{f}_{1}\acute{f}_{1}}{f_{1}f_{1}}
+\frac{2\acute{f}_{1}\acute{f}_{2}}{f_{1}f_{2}}-\frac{N^{2}f^{2}_{2}}{4f^{4}_{1}}.\label{12aMar2012}
\end{equation}
Here we emphasize that (\ref{12aMar2012}) plays a role to connect
the equivalence between the Hamiltonian constraint ${\cal{H}}$ and
Einstein equation $G_{00}$, i.e., ${\cal{H}}\sim{}G_{00}$. Such
equivalence relation is not only for Bianchi II model, but also
true for all cases. We will explicate this relation in section 4.

On the other hand, we compute the second part of the
Euler-Lagrange equation~\cite{RWR}
\begin{eqnarray}
\frac{\delta{}{\cal{L}}}{\delta{}g_{ij}}
=-\sqrt{g}N(R^{ij}-g^{ij}R)
-\frac{2N}{\sqrt{g}}\left(p^{im}p^{j}{}_{m}-\frac{1}{2}pp^{ij}\right)
+\sqrt{g}G^{ijmn}\nabla_{m}\nabla_{n}N+\pounds_{\xi}p^{ij},
\label{20aJuly2011}
\end{eqnarray}
One can tune $\lambda$ in such a way that the universe has the
same expanding rate such that $\xi=0$. The corresponding non-zero
components are
\begin{eqnarray}
&&\frac{\delta{\cal{L}}}{\delta{}g_{11}}=-\frac{2p^{11}\acute{f}_{1}}{f_{1}},\quad
\frac{\delta{\cal{L}}}{\delta{}g_{22}}=-2\left(p^{22}-\frac{p}{f^{2}_{2}}\right)\frac{\acute{f}_{1}}{f_{1}},\\
&&\frac{\delta{\cal{L}}}{\delta{}g_{23}}=-\frac{2p^{23}\acute{f}_{1}}{f_{1}},\quad
\frac{\delta{\cal{L}}}{\delta{}g_{33}}=-\frac{2p^{33}\acute{f}_{1}}{f_{1}}.
\end{eqnarray}
Consequently, explicitly list out the Euler-Lagrange equations as
follows
\begin{eqnarray}
&&0=\frac{\partial{}p^{11}}{\partial\lambda}-\frac{\delta{\cal{L}}}{\delta{}g_{11}}
=-g^{11}\frac{\partial}{\partial\lambda}\left[\frac{\sqrt{g}}{N}\left(\frac{\acute{f}_{1}}{f_{1}}
+\frac{\acute{f}_{2}}{f_{2}}\right)\right],\\
&&0=\frac{\partial{}p^{22}}{\partial\lambda}-\frac{\delta{\cal{L}}}{\delta{}g_{22}}
=\frac{1}{f_{1}}\left(\frac{2\acute{f}_{1}}{f_{2}\acute{f}_{2}}-\frac{x^{2}}{f_{1}}\right)
\frac{\partial}{\partial\lambda}\left[\frac{\sqrt{g}}{N}\left(\frac{\acute{f}_{1}}{f_{1}}
+\frac{\acute{f}_{2}}{f_{2}}\right)\right],\label{9aMar2012}\\
&&0=\frac{dp^{23}}{d\lambda}-\frac{\delta{\cal{L}}}{\delta{}g_{23}}
=-g^{23}\frac{\partial}{\partial\lambda}\left[\frac{\sqrt{g}}{N}\left(\frac{\acute{f}_{1}}{f_{1}}
+\frac{\acute{f}_{2}}{f_{2}}\right)\right],\\
&&0=\frac{\partial{}p^{33}}{\partial\lambda}-\frac{\delta{\cal{L}}}{\delta{}g_{33}}
=-g^{33}\frac{\partial}{\partial\lambda}\left[\frac{\sqrt{g}}{N}\left(\frac{\acute{f}_{1}}{f_{1}}
+\frac{\acute{f}_{2}}{f_{2}}\right)\right].
\end{eqnarray}
Thus, we have the general result
\begin{eqnarray}
0=\frac{\partial}{\partial\lambda}\left[\frac{\sqrt{g}}{N}
\left(\frac{\acute{f}_{1}}{f_{1}}+\frac{\acute{f}_{2}}{f_{2}}\right)\right],\label{27cNov2011}
\end{eqnarray}
provided that $f_{2}$ cannot be a constant which has already
exhibited in (\ref{9aMar2012}), i.e., $\acute{f}_{2}$ at the
denominator. Basically, there are only two equations indicated in
(\ref{12aMar2012}) and (\ref{27cNov2011}) which is exactly
matching with the two unknown functions. While GR gives 3
equations and 2 unknowns, forming a completeness for finding the
solutions, we confused that why we have more equations than we
expected empirically. Here we compare the results between 3-space
and 4-metric, remembering that $dt=Nd\lambda$, rewrite
(\ref{27cNov2011}) in terms of $G_{\mu\nu}$
\begin{eqnarray}
0&=&\frac{d}{d\lambda}\left[f_{1}\frac{d}{d\lambda}(f_{1}f_{2})\right]\nonumber\\
&=&\sqrt{g}(G_{00}+f^{-2}_{1}G_{11}),
\end{eqnarray}
under the circumstance that $f_{2}$ is not real.  This means that,
if we know this particular restriction in 4D, we only need
$G_{00}$ and $G_{11}$, while $G_{22}$ becomes not necessary.
However, without $G_{22}$, constant $f_{2}$ is allowable for
$G_{00}$ and $G_{11}$ as mentioned in section 2, i.e.,
$(f_{1},f_{2})=(\sqrt{t},1)$. Based on the above discussion, the
3-space approach cannot practise any advantage than 4-metric to
treat the same problem, but the importance is that it seems really
showing a more fundamental concept as Dirac
suggested~\cite{Dirac}.

Searching a relationship between $f_{1}$ and $f_{2}$. Consider
(\ref{27cNov2011}) and let the function inside the square bracket
be a constant $k$, i.e.,
\begin{equation}
f_{2}=\frac{1}{f_{1}}\int\frac{Nk}{f_{1}}d\lambda.
\end{equation}
This shows $(f_{1},f_{2})$ couple together. Generally, given
$f_{2}$ and then $f_{1}$ could be solved. But the question is how
to select $f_{2}$ in a systematical way such that both
$(f_{1},f_{2})$ satisfy (\ref{12aMar2012}) and (\ref{27cNov2011})
simultaneously? We are more favourable the 3-space theory instead
of 4-metric. Not only the concept is more basic, but also the
mathematical manipulation is a bit easier to solve by selecting a
specific $N$. Consider (\ref{27cNov2011}) again
\begin{equation}
\frac{\sqrt{g}}{N}\left(\frac{\acute{f}_{1}}{f_{1}}+\frac{\acute{f}_{2}}{f_{2}}\right)=k.\label{7aMar2013}
\end{equation}
Comparing (\ref{12aMar2012}) and (\ref{7aMar2013}), one may obtain
\begin{equation}
\frac{\acute{f}_{2}\acute{f}_{2}}{f_{2}f_{2}}
+\frac{N^{2}}{4f^{4}_{1}}\left(f^{2}_{2}-\frac{4k^{2}}{f^{2}_{2}}\right)=0.
\end{equation}
In order to decouple $(f_{1},f_{2})$, simply allowing $k=1/2$ and
seeking a suitable $N$. The general solutions are
\begin{eqnarray}
f_{2}=\frac{A_{M}}{(\cosh\lambda)^{1/2^{M}}},\quad{}
f_{1}=\frac{B_{M}}{f_{2}}\exp\int\frac{1}{\sqrt{1+f^{2}_{2}}}\prod^{M}_{n=1}\frac{1}{2}\sqrt{1+f^{2^{n}}_{2}}d\lambda,
\end{eqnarray}
where $M=1,2,3,...$.  We find that $A_{M}=\pm1$ for all $M$,
$B_{M}$ are constants and
\begin{equation}
N=\frac{2\sqrt{g}}{\sqrt{1+f^{2}_{2}}}\prod^{M}_{n=1}\frac{1}{2}\sqrt{1+f^{2^{n}}_{2}}.
\end{equation}
For example, when $M=1$ which means choosing $N=\sqrt{g}$, we
recover the results $(f_{1},f_{2})$ in~\cite{Singh} as indicated
in (\ref{29aAug2012}).

Furthermore, using BSW method one more time, we find another set
of solutions
\begin{equation}
f_{2}=A_{M}(\sin{\lambda})^{1/2^{M}}, \quad{}
f_{1}=\frac{B_{M}}{f_{2}}\exp\int\frac{N}{2\sqrt{g}}d\lambda,
\end{equation}
again $M$ is a positive integer, all $A_{M}=\pm1$, $B_{M}$ are
real and
\begin{equation}
N=\frac{2f^{2}_{1}}{\sqrt{1+f^{2}_{2}}}\prod^{M}_{n=1}\frac{\sqrt{1+f^{2^{n}}_{2}}}{2f^{2^{(n-1)}}_{2}},
\end{equation}
In particular, we explicitly write out the first two solutions
\begin{eqnarray}
&&M=1,\quad{}f_{2}=\pm\sqrt{\sin\lambda},\quad\quad{}f_{1}=\frac{B_{1}}{\sqrt{1+\cos\lambda}},\label{3bSep2012}\\
&&M=2,\quad{}f_{2}=\pm(\sin\lambda)^{1/4},\quad{}f_{1}=\frac{B_{2}}{f_{2}}
\left[\frac{\sqrt{\sin{\lambda}}}{\sqrt{1-\sin{\lambda}}+1}\right]^{1/2},
\end{eqnarray}
provided that $N_{1}=f^{2}_{1}/f_{2}$,
$N_{2}=f^{2}_{1}\sqrt{1+f^{4}_{2}}/(2f^{3}_{2})$ and
$\lambda\in(0,\pi/2)$. Footnote: here we remind the reader that
$(N,f_{1},f_{2})=(f^{2}_{1}/f_{2},1/\sqrt{1-\sin\lambda},\sqrt{\cos\lambda})$
is not a new solution since it is duplicated with
(\ref{3bSep2012}) by a simple transformation:
$\lambda\rightarrow\lambda-\pi/2$.

\section{Hamiltonian initial value constraint}
Here we try to reproduce (\ref{12aMar2012}) using the Hamiltonian
initial value constraint
\begin{eqnarray}
{\cal{H}}&=&\frac{1}{\sqrt{g}}G_{abcd}p^{ab}p^{cd}-{}^{3}R\sqrt{g}\nonumber\\
&=&-\frac{2\sqrt{g}}{N^{2}}\left(\frac{\acute{f}_{1}\acute{f}_{1}}{f_{1}f_{1}}
+\frac{2\acute{f}_{1}\acute{f}_{2}}{f_{1}f_{2}}-\frac{N^{2}f^{2}_{2}}{4f^{4}_{1}}\right)\nonumber\\
&=&-2\sqrt{g}G_{00},\label{14aMay2013}
\end{eqnarray}
where $G_{abcd}=g_{ac}g_{bd}-\frac{1}{2}g_{ab}g_{cd}$. Therefore,
under the restriction of the Bianchi II model, we deduce that
${\cal{H}}\sim{}G_{00}$. But, looking at (\ref{14aMay2013})
carefully, one may suspect that whether this is a special result
for this specific cosmological model?  It is not likely that
satisfy the other models. Since ${\cal{H}}$ only contain the
non-spatial derivative up to $\acute{g}_{ab}$, while $G_{00}$
should include second time derivative of $g_{\mu\nu}$ in general.
Here we claim that ${\cal{H}}\sim{}G_{00}$ is valid in principle.
The verification is follows. Having removed away
$\ddot{g}_{\mu\nu}$ by redefine the time coordinate in such a way
that $g_{0c}=0$, we find
\begin{eqnarray}
2G_{00}=-\frac{1}{4}G^{abcd}\dot{g}_{ab}\dot{g}_{cd}-{}^{3}Rg_{00}.
\end{eqnarray}
On the other hand, using the following identity
\begin{eqnarray}
\frac{\partial{}g_{ab}}{\partial\lambda}=\frac{2N}{\sqrt{g}}G_{abcd}p^{cd}+\pounds_{\xi}g_{ab}.
\end{eqnarray}
Consider the Hamiltonian constraint once more in different symbols
\begin{eqnarray}
{\cal{H}}&=&\sqrt{g}\left[\frac{1}{4N^{2}}G^{abcd}(\acute{g}_{ab}-\pounds_{\xi}g_{ab})(\acute{g}_{cd}
-\pounds_{\xi}g_{cd})-{}^{3}R\right]\nonumber\\
&=&-2\sqrt{g}G_{00},\label{10bMay2013}
\end{eqnarray}
provided that $g_{00}=-1$ which is legitimate by redefining the
time coordinate again.

\section{Conclusion}
Baierlein, Sharp and Wheeler proposed a Jacobi type timeless
Lagrangian based on the 3-dimensional geometry of space to
reproduce the same physics as the 4-metric does.  Barbour $et$.
$al$. further extended this idea and they name it 3-space
approach. We think the concept of 3-space idea is more fundamental
than 4-metric GR as Dirac pointed out 55 years ago. Here we use
Bianchi II cosmological model as an example to illustrate this
3-space idea. We examine that the Hamiltonian constraint
${\cal{H}}$ is equivalent to Einstein equation $G_{00}$ in
general. Moreover, We find that it is a bit easier to recover the
known result which satisfy the Einstein equation in vacuum, on the
other hand, we also find another new set of solutions.


\begin{thebibliography}{3}

\bibitem{Dirac}
Dirac P A M 1958 {\it{}Proc. Roy. Soc. A} $\mathbf{246}$ 333

\bibitem{BSW}
Baierlein R F, Sharp D and Wheeler J A 1962 {\it{Phys. Rev.}}
$\mathbf{126}$ 1864

\bibitem{RWR}
Barbour J, Foster B Z and $\acute{\mbox{O}}$ Murchadha N 2002
{\it{Class. Quantum Grav.}} $\mathbf{19}$ 3217

\bibitem{Singh}
Ram S and Singh C P 1998 {\it{}Astrophys. Space Sci.} {\bf{}257}
287

\bibitem{metricII}
Alex Harvey 1983 {\it Phys. Rev.} D {\bf 28} 2121


\end{thebibliography}
\end{document}